\documentclass[aps,floatfix]{revtex4}
\usepackage{mathrsfs}
\usepackage{amssymb}
\usepackage{amsmath}
\usepackage{graphicx}
\usepackage[normalem]{ulem}
\usepackage[dvips]{color}
\usepackage{bm}
\usepackage{longtable}

\renewcommand\sout{\bgroup \color{red} \ULdepth=-.5ex \ULset}

\renewcommand{\rm}[1]{\textrm{#1}}

\def\esym{$E_{sym}(\rho)$~}

\def\es0{$E_{sym}(\rho_0)$}

\def\us0{$U_{sym}(\rho_0,k_F)$~}

\def\lr{$L(\rho)$~}

\def\l0{$L(\rho_0)$~}

\renewcommand{\rm}[1]{\textrm{#1}}

\begin{document}

\title{Nuclear Symmetry Energy Extracted from Laboratory Experiments}

\author{Bao-An Li\footnote{Email: Bao-An.Li@Tamuc.edu}}
\affiliation{Department of Physics and Astronomy, Texas A$\&$M University-Commerce, Commerce, TX 75429-3011, USA}

\maketitle

\noindent{\bf Introduction}\\
The Equation of State (EOS) of uniform neutron-rich nucleonic matter of isospin asymmetry $\delta=(\rho_n-\rho_p)/\rho$ and density $\rho$ can be expressed in terms of the energy per nucleon $ E(\rho ,\delta )$ within the parabolic approximation as
\begin{equation}\label{eos1}
E(\rho ,\delta )=E(\rho,0)+E_{\rm sym}(\rho )\cdot \delta ^{2} +\mathcal{O}(\delta^4)
\end{equation}
where $E_{\rm sym}(\rho )= \frac{1}{2}[\partial ^{2}E(\rho,\delta )/\partial \delta ^{2}]_{\delta =0}\approx E(\rho,1)-E(\rho,0)$
is the symmetry energy of asymmetric nuclear matter (ANM). It is approximately the energy cost of converting symmetric nuclear matter (SNM, with equal numbers of protons and neutrons) into pure neutron matter (PNM).
Many interesting questions including the dynamics of supernova explosions, heavy-ion collisions, structures of neutron stars and rare isotopes, frequencies and strain amplitudes of gravitational waves from both isolated pulsars and collisions involving neutron stars all depend critically on the EOS of neutron-rich nucleonic matter. Thanks to the great efforts of scientists in both nuclear physics and astrophysics over the last four decades, much knowledge about the EOS of SNM, i.e, the $E(\rho,0)$ term in Eq. (\ref{eos1}), has been obtained \cite{dan02}. In more recent years, significant efforts have been devoted to exploring the poorly known \esym using both terrestrial laboratory experiments and astrophysical observations \cite{ireview98,ibook01,Steiner05,ditoro,LCK08,Lynch09,Trau12,Tsang12,Lat13,Tesym,Chuck14,Baldo16}. Theoretically, essentially all available nuclear forces have been used to calculate the \esym within various microscopic many-body theories and/or phenomenological models. However, model predictions still vary largely at both sub-saturation and supra-saturation densities although they agree often by construction at the saturation density $\rho_0$. Therefore, accurate experimental constraints are imperative for making further progresses in our understanding of the density dependence of nuclear symmetry energy. To facilitate the extraction of information about the \esym from laboratory experiments, much work has been done to find observables that are sensitive to the \esym by studying static properties, excitations and collective motions of nuclei as well as various observables of nuclear reactions.  Comprehensive reviews on the recent progress and remaining challenges in constraining the \esym can be found in the literature \cite{ireview98,ibook01,Steiner05,ditoro,LCK08,Lynch09,Trau12,Tsang12,Lat13,Tesym,Chuck14,Baldo16,Blaschke16}. Most importantly, much progress has been made in constraining the \esym around and below the saturation density $\rho_0$ while its high density behavior remains rather uncertain. Combining results from ongoing and planned new laboratory experiments with radioactive beams and astrophysical observations using advanced x-ray observatories and gravitational wave detectors has the great promise of determining the symmetry energy of dense neutron-rich matter in the near future. \\

\noindent{\bf Physics underlying nuclear symmetry energy}\\
It is well known that the nucleon potential $U_{n/p}(k,\rho,\delta)$ in ANM can be expanded up to the second order in $\delta$ as
\vspace{-0.1cm}
\begin{equation}\label{sp}
U_{\tau}(k,\rho,\delta)=U_0(k,\rho)+\tau_3 U_{sym,1}(k,\rho)\cdot\delta+U_{sym,2}(k,\rho)\cdot\delta^2+\mathcal{O}(\delta^3),
\end{equation}
where $\tau_3=\pm$ for $\tau=n/p$ and $U_0(k,\rho)$, $U_{sym,1}(k,\rho)$ and $U_{sym,2}(k,\rho)$ are the isoscalar, isovector (symmetry or Lane potential~\cite{Lan62}) and the second-order isoscalar potentials, respectively.
At the mean-field level, using the Bruckner theory \cite{bru64,Dab73} or the Hugenholtz-Van Hove (HVH) theorem \cite{hug}, the \esym and its density slope 
$L(\rho) \equiv \left[3 \rho (\partial E_{\rm sym}/\partial \rho\right)]_{\rho}$ at an arbitrary density $\rho$ can be expressed generally as \cite{XuLiChen10a,xuli2,Rchen}
\vspace{-0.1cm}
\begin{eqnarray}
E_{\rm sym}(\rho) &=&\frac{1}{3} \frac{\hbar^2 k_F^2}{2 m_0^*} +
\frac{1}{2} U_{\rm sym,1}(\rho,k_{F}), \label{Esymexp2}
\\
L(\rho) &=& \frac{2}{3} \frac{\hbar^2 k_F^2}{2 m_0^*} + \frac{3}{2} U_{\rm sym,1}(\rho,k_F) 
- \frac{1}{6}\Big(\frac{\hbar^2 k^3}{{m_0^*}^2}\frac{\partial m_0^*}{\partial k} \Big)|_{k_F} 
+\frac{dU_{\rm sym,1}}{dk}|_{k_F} k_F 
+ 3U_{\rm sym,2}(\rho,k_F), \label{Lexp2}
\end{eqnarray}
where $k_F=(3\pi^2\rho/2)^{1/3}$ is the nucleon Fermi momentum and $m^*_0/m=(1+\frac{m}{\hbar^2k_{\rm F}}dU_0/dk)^{-1}|_{k_F}$ is the nucleon isoscalar effective mass. 
The \esym consists of two terms, i.e., the kinetic symmetry energy $E^1_{sym}(\rho)$ equivalent to $1/3$ the Fermi energy of quasi-nucleons with an isoscalar effective mass $m^*_0$ and the potential symmetry energy $E^2_{sym}(\rho)$ equivalent to $1/2$ the isovector potential $U_{\rm sym,1}(\rho,k_{F})$ at $k_F$. The density slope $L(\rho)$ can be cast into five terms depending on (1) the isoscalar effective mass $m^*_0$ (defined as $L_1(\rho)\equiv\frac{2 \hbar^2 k_{\rm F}^2}{6 m_0^{\ast}(\rho,k_{\rm F})}$), (2) the momentum dependence of $m^*_0$ ($L_2(\rho)\equiv-\frac{ \hbar^2 k_{\rm F}^3}{6 m_0^{\ast 2}(\rho,k_{\rm F})} \frac{\partial m_0^{\ast}(\rho, k)}{\partial k}|_{k = k_{\rm F}}$), (3) the isovector potential $U_{\rm sym,1}(\rho,k_F)$ ($L_3(\rho)\equiv\frac{3}{2} U_{\rm sym, 1}(\rho, k_{\rm F})$), (4) the momentum dependence of the isovector potential ($L_4(\rho) \equiv \frac{\partial U_{\rm sym, 1}(\rho, k)}{\partial k}|_{k = k_{\rm F}} \cdot k_{\rm F}$) and (5) the second-order isoscalar potential $U_{\rm sym,2}(\rho,k_F)$ ($L_5(\rho) \equiv 3 U_{\rm sym, 2}(\rho, k_{\rm F})$). 
A relativistic version of the decomposition of the $E_{sym}(\rho)$ and $L(\rho)$ in terms of the Lorentz covariant nucleon self-energies can be found in ref. \cite{Cai-EL}.

Obviously, the \esym and its density slope \lr depend on the density and momentum dependence of both the isoscalar and isovector potentials. While the $U_0(k,\rho)$ has been relatively well constrained by studying experimental observables in heavy-ion reactions, especially various kinds of collective flow and kaon production, our current knowledge about the density and momentum dependence of the isovector potential $U_{sym,1}(k,\rho)$ and the second-order isoscalar potential $U_{\rm sym,2}(\rho,k)$ is very poor. In the literature, some calculations consider the momentum dependence of both the isoscalar and isovector potentials due to the finite-range parts of nuclear interactions, while others consider only the momentum dependence of one or none of them, leading to very model-dependent predictions for the \esym and $L(\rho)$. On the other hand, different experimental observables may be sensitive to different components of the \esym and $L(\rho)$, providing multiple probes. However, this feature may also lead to possible variations of the results extracted from different experiments as the data analyses often rely on models which may or may not consider all parts of the \esym and $L(\rho)$. It is worth emphasizing that while the \esym depends only on the magnitudes of $m^*_0$ and $U_{\rm sym,1}$ at $k_F$, extra quantities characterizing the momentum dependence of both the $m^*_0$ and $U_{\rm sym,1}$ as well as the magnitude of $U_{\rm sym,2}$ at $k_F$ are required to determine completely the $L(\rho)$. Of course, as indicated in Eqs. 3 and 4, the $E_{\rm sym}(\rho)$ and \lr are intrinsically correlated. However, since the \lr depends on three more quantities that are poorly known, their correlation is not unique and the \lr is more uncertain as demonstrated by various model analyses of sometimes the same data. 

It is important to point out that the \esym is closely related to the neutron-proton effective mass splitting $m^*_{n-p}\equiv(m_{\rm n}^*-m_{\rm p}^*)/m$, which is a fundamental quantity having broad impacts on many interesting issues in both nuclear physics and astrophysics \cite{LiChen16,LiChen16b}. In terms of the momentum dependence of the single-nucleon potential or the \esym and $L(\rho)$, one has
\begin{equation}\label{npe1}
m^*_{n-p}\approx\frac{2\delta m}{\hbar^2k_F}\left[-\frac{dU_{sym,1}}{dk}-\frac{k_F}{3}\frac{d^2U_0}{dk^2}+\frac{1}{3}\frac{dU_0}{dk}\right]\left(\frac{m^*_0}{m}\right)^2
\approx\frac{\delta}{E_F(\rho)}\left[3E_{\rm sym}(\rho)-L(\rho)-\frac{1}{3}\frac{m}{m^*_0}E_{F}(\rho)\right]\left(\frac{m_0^*}{m}\right)^2\nonumber
\end{equation}
where $E_F(\rho)$ is the Fermi energy in SNM at density $\rho$. Therefore, while probing the density dependence of the nuclear symmetry energy, we are also studying the neutron-proton effective mass splitting in neutron-rich nuclear matter \cite{LiChen16,LiChen16b,Lynch16}. 

It is well known that the kinetic symmetry energy is due to the Pauli blocking and the different Fermi momenta of quasi-nucleons. Since the nucleon isoscalar effective mass $m^*_0/m\approx 0.7$ at $\rho_0$, the kinetic symmetry energy of quasi-nucleons at $\rho_0$ is about 43\% larger than that of the free Fermi gas of about 12 MeV frequently used in the literature. The potential symmetry energy is due to the isospin dependence of the strong interaction. For example, the Hartree term of the isovector potential at $k_F$ in the interacting Fermi gas model can be written as \cite{pre,Xu-tensor} 
$
U_{sym,1}(k_F,\rho)=
\frac{1}{4}\rho\int [V_{T1}(r_{ij})f^{T1}(r_{ij})-V_{T0}(r_{ij})f^{T0}(r_{ij})]d^3r_{ij}
$
in terms of the isosinglet (T=0) and isotriplet (T=1) nucleon-nucleon (NN) interactions $V_{T0}(r_{ij})$ and $V_{T1}(r_{ij})$, and the corresponding NN correlation functions $f^{T0}(r_{ij})$ and $f^{T1}(r_{ij})$, respectively. 
While the charge independence of the strong interaction requires that $V_{nn}=V_{pp}=V_{np}$ in the T=1 channel, they are not necessarily equal to the $V'_{np}$ in the T=0 channel. 
Obviously, if there is no isospin dependence in both the NN interaction and the correlation function, then the isovector potential $U_{sym,1}(k_F,\rho)$ vanishes. The momentum dependence of the 
isovector potential from the Fock term using Gogny-type finite-range, isospin-dependent interactions \cite {Gogny} is often parameterized by using different strengths of interactions between like and unlike nucleon pairs \cite{Das03}. Indeed, microscopic many-body theories \cite{Bom91,Die03} predicted that the potential symmetry energy is dominated by the isosinglet interaction channel and the contribution from the isotriplet channel almost vanishes. It is also well known that the short range correlation in the T=0 channel is much stronger than that in the T=1 channel \cite{bethe,Sub08,Hen14}. The potential symmetry energy thus reflects the 
isospin dependence of nucleon-nucleon interactions and correlations. 
\begin{figure}[h]
\centering
\includegraphics[width=10cm,height=6cm]{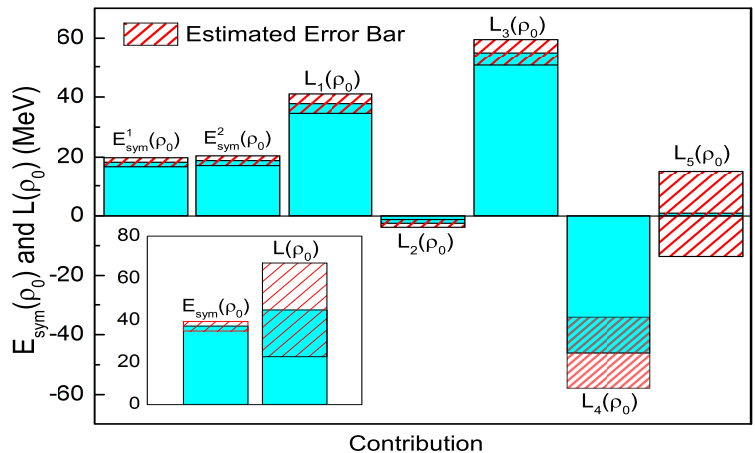}
\caption{The kinetic $E^1_{sym}(\rho_0)$ and potential $E^2_{sym}(\rho_0)$ contributions to the symmetry energy $E_{\mathrm{sym}}(\rho_0)$ and the five components of the slope parameter $L(\rho_0)$ at saturation density extracted from nucleon-nucleus elastic scattering data.  Taken from ref.\ \cite{LiXH13}.}
\label{EsymL}
\end{figure}

The momentum dependence of the isoscalar and isovector potential at $\rho_0$ has been explored extensively using (p,n) charge-exchange and nucleon-nucleus elastic scatterings. The resulting single-nucleon potential has been used to constrain the \es0 and $L(\rho_0)$ \cite{XuLiChen10a,LiXH13}. As an example, shown in Fig. \ref{EsymL} are the kinetic ($E^1_{sym}(\rho_0)$) and potential ($E^2_{sym}(\rho_0)$) parts of the symmetry energy \es0 and the five components of its slope $L(\rho_0)$ extracted from a recent optical model analyses of the nucleon-nucleus elastic scattering data \cite{LiXH13}. The kinetic and potential parts of the symmetry energy from this analysis are both approximately equal to $18$ MeV. Among the five parts of the slope parameter $L(\rho_0)$, the $L_4$ due to the momentum-dependence of the symmetry potential and the $L_5$ from the second-order isoscalar potential have the largest uncertainties. The characteristically decreasing isovector potential with increasing energy/momentum leads to a positive neutron-proton effective mass splitting \cite{Li-han} and a negative value of $L_4$ at $\rho_0$. The mere fact that \lr has five terms having different signs and physical origins indicates clearly the difficulties of completely pinning it down even at the saturation density.\\

Given our incomplete knowledge about some components of the $L(\rho)$ as we discussed above, no wonder why it is so difficult to determine accurately the incompressibility of ANM at $\rho_0$
\begin{equation}
K(\delta)= K_0+K_{\tau}\delta^2+\mathcal{O}(\delta^4)
\end{equation}
where $K_{\tau}=K_{\rm{sym}}-6L(\rho_0)-\frac{Q_0L(\rho_0)}{K_0}$ in terms of the curvature of the symmetry energy
\begin{equation}
K_{\rm{sym}}\equiv\left[9\rho^2\frac{\partial^2E_{\rm{sym}}(\rho)}{\partial\rho^2}\right]_{\rho_0}
=3\left[\rho\frac{\partial L(\rho)}{\partial\rho}-L(\rho)\right]_{\rho_0},
\end{equation}
as well as the skewness $Q_0\equiv27\rho_0^3\partial^3E(\rho,0)/\partial\rho^3|_{\rho=\rho_0}$ and incompressibility $K_0$ of SNM at $\rho_0$.
As the $K_{\rm{sym}}$ involves the derivative $\frac{\partial L(\rho)}{\partial\rho}$, to determine its value we have to know not only the magnitudes of $\frac{dU_{\rm sym,1}}{dk}$ and $U_{sym,2}$ 
but also their density and momentum dependences. Unfortunately, these quantities are largely unknown both theoretically and experimentally. As a result, the current estimate of $K_{\tau}\approx -550\pm 100$ MeV \cite{Colo1} from analyzing many different kinds of experimental data available has a large error bar.  It is also not surprising that some of the best models available are having troubles to reproduce the incompressibility of some neutron-rich nuclei, e.g., Tin isotopes from $^{112}$Sn to $^{124}$Sn \cite{TLi,Jorge-TLi}.

The decomposition of \esym and \lr in Eqs. 3 and 4 according to the density and momentum dependence of single-nucleon potentials at the mean-field level is transparent and useful for identifying the important underlying physics. We emphasize, however, it has its limitations. There are density regions or phenomena for which correlations beyond the mean-field level have to be treated properly. For example, effects of the tensor force on the \esym are averaged out at the mean-field level. In fact, effects of the tensor force on the \esym have been a longstanding issue \cite{kuo65,Eng98,Lee2,Car13}. In particular, 
tensor force induced short-range correlations may alter significantly the relative contributions of the kinetic and potential parts to the total symmetry energy \esym and its slope \lr\cite{Xulili,Vid11,Car12,Hen15b,Cai-d4,Cai-RMF}.
While how the total symmetry energy is divided into kinetic and potential parts seems to have no obvious effect on describing properties of neutron stars as it is the total pressure and energy density that are needed in solving the 
Tolman-Oppenheimer-Volkoff (TOV) equation \cite{Hen-Steiner}, it is certainly important for simulating nuclear reactions using transport models describing the evolution of quasi-nucleons in phase space under the influence of 
nuclear mean-fields and collision integrals \cite{LiG15,Yong-PLB}. The strong isospin dependence of the tensor force may even lead to negative \esym at high densities \cite{Pan72,Wir88a}, leading to the prediction of some interesting new phenomena in neutron stars \cite{Kut93,Kut94,Kut06}. Indeed, going beyond the mean-field level, various microscopic many-body theories which incorporate correlations to differing degrees have been used to predict the density dependence of the symmetry energy. Unfortunately, considering the diverse features of various microscopic model studies using different interactions currently available in the literature, the predictions diverge broadly at supra-saturation densities, and it is not always clear what underlying physics is responsible for a particular feature of a prediction. 

The EOS of uniform and isospin-asymmetric nucleonic matter described by Eq. 1 and the definition of its symmetry energy have their ranges of validity. For example, at low densities below the so-called Mott points, various clusters start forming. One thus has to go beyond the mean-field by considering correlations/fluctuations and in-medium properties of clusters in constructing the EOS of stellar matter for astrophysical applications
\cite{Lat-Eos,Shen,Hor06,Joe10,Hag12,Rop13,Typ14,Hag14}. Then the Eq. 1 is obviously no longer valid and there seems to be no need to introduce a symmetry energy of clustered matter for describing its EOS. In fact, for the clustered matter, because of the different binding energies of mirror nuclei, Coulomb interactions, different locations of proton and neutron drip lines in the atomic chart, the system no longer possesses a proton-neutron exchange symmetry. Moreover, different clusters in the medium have their own local internal isospin asymmetries and densities. Indeed, in terms of the average density $\rho_{av}$ and the average isospin asymmetry $\delta_{av}$ of the whole system, the EOS of clustered matter has been found to have odd terms in $\delta_{av}$ that are appreciable compared to the $\delta^2_{av}$ term \cite{Agr14,Fan14}. Thus, it is conceptually ambiguous to define a symmetry energy for clustered matter in the same sense as for the uniform nucleonic matter. Nonetheless, either the second-order derivative of energy per nucleon $e_{\rm{cluster}}(\rho_{av},\delta_{av})$ in clustered matter with respect to $\delta_{av}$, i.e., $E^{\rm cluster}_{\rm sym}(\rho_{av})\equiv\frac{1}{2}[\partial ^{2}e_{\rm cluster}/\partial \delta_{av} ^{2}]_{\delta_{av} =0}$, 
or the quantity $E^{\rm cluster}_{\rm sym}(\rho_{av})\equiv 1/2[e_{\rm{cluster}}(\delta_{av}=1)+ e_{\rm{cluster}}(\delta_{av}=-1)-2e_{\rm{cluster}}(\delta_{av}=0)]$ as if the EOS is parabolic in $\delta_{av}$, has been used in studying the symmetry energy of clustered matter. This quantity stays finite at the limit of zero average density. However, it is important to stress here that the $E^{\rm cluster}_{\rm sym}(\rho_{av})$ has physical meanings very different from those of the \esym for uniform nucleonic matter in Eq. 1. \\
 
\noindent{\bf Symmetry energy at the saturation density}\\
It is customary to characterize the density dependence of nuclear symmetry energy near $\rho_0$ by using the $E_{\rm sym}(\rho_0)$ and $L(\rho_0)$. 
In recent years, much progress has been made in constraining them using various observables from both terrestrial laboratory experiments and astrophysical observations.
Depending on the observables and models used, various correlations between $E_{\rm sym}(\rho_0)$ and $L(\rho_0)$ have been obtained.
Interestingly, most of them overlap around $E_{sym}(\rho_0)\sim 31$ MeV and $L(\rho_0)\sim 55$ MeV according to ref. \cite{ste14} using selected 6 analyses. 
\begin{figure}[htb]
\includegraphics[width=18cm,height=5cm]{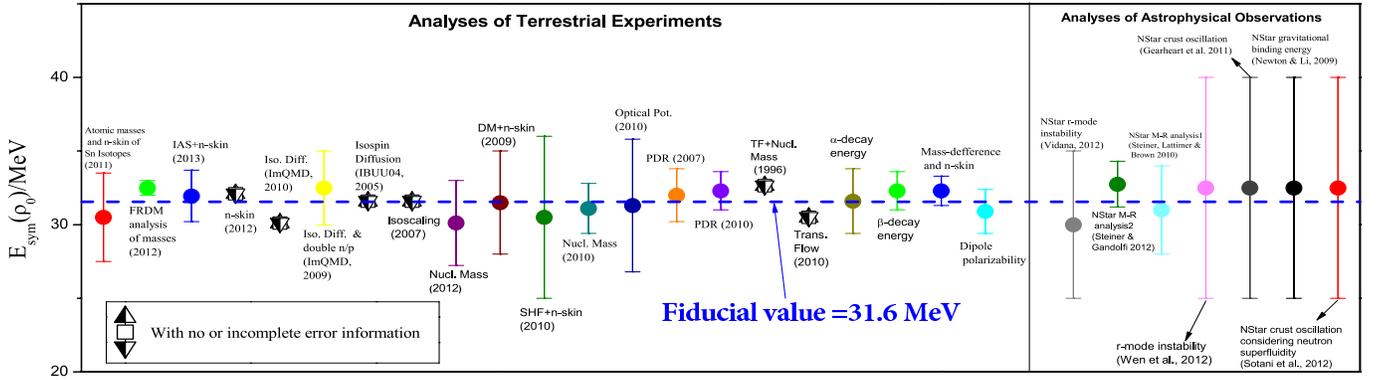}

\caption{{\protect Central values of \es0 from 28 model analyses of terrestrial nuclear experiments and astrophysical observations. Taken from ref. \cite{Li-han}.
}}\label{Esym0-Li}
\end{figure}
\begin{figure}[htb]
\includegraphics[width=18cm,height=5cm]{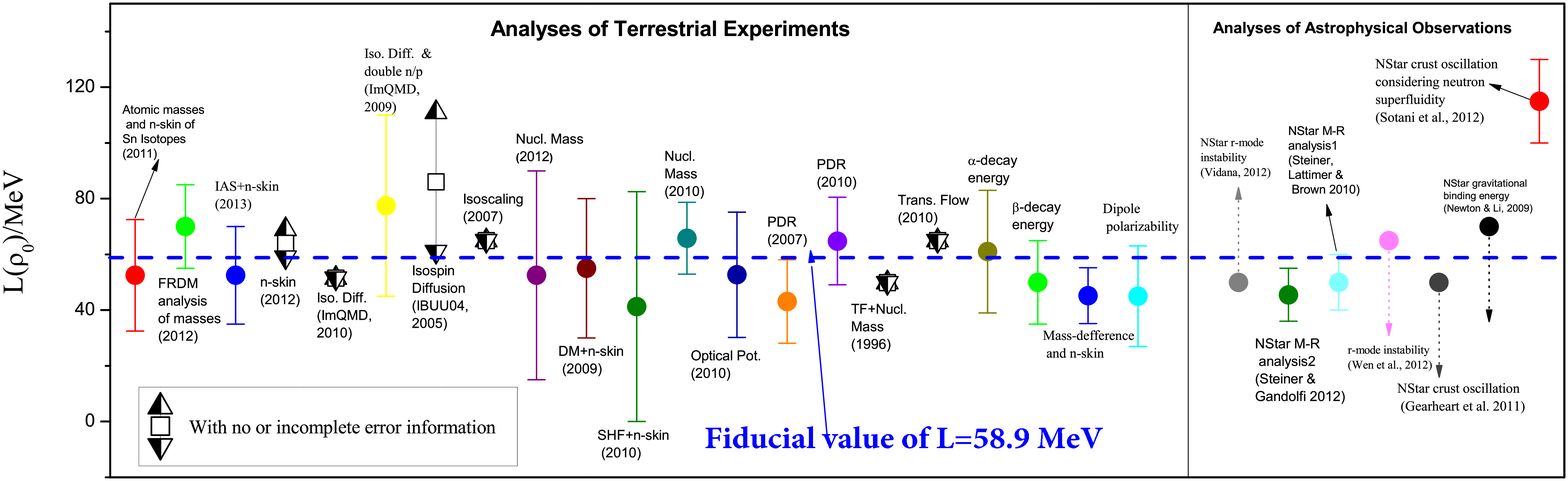}
\caption{{\protect Central values of $L(\rho_0)$ from the same analyses as in Fig. \ref{Esym0-Li}. Taken from ref. \cite{Li-han}.
}}\label{L0-Li}
\end{figure}

Within their respective uncertainties or ranges of validity, as shown in Figs. 2 and 3, a survey of 28 analyses made before August, 2013 \cite{Li-han} found that the central values of the $E_{\rm sym}(\rho_0)$ and $L(\rho_0)$ scatter around $31.6\pm 2.66$ MeV and $58.9\pm 16$ MeV, respectively. Observables used in these analyses include the atomic masses, neutron-skins of heavy nuclei, isospin diffusion in heavy-ion reactions, excitation energies of isobaric analog states (IAS), isoscaling of fragments from intermediate energy heavy-ion collisions, the electric dipole polarizability from analyzing the Pygmy dipole resonance, the frequency of isovector giant dipole resonances, $\alpha$ and $\beta$ decay energies, optical potentials from analyzing nucleon-nucleus scatterings, and several observables of neutron stars, etc. Very recently, a more extensive survey of 53 analyses up to October, 2016 found that the central values of $E_{\rm sym}(\rho_0)$ and $L(\rho_0)$ are $31.7\pm 3.2$ MeV and $58.7\pm 28.1$ MeV \cite{Oer16}, respectively. Clearly, these results are consistent with those from the earlier surveys but with slightly larger uncertainties when a more diverse set of data and models are used. Interestingly, all three surveys found empirically that $L(\rho_0) \approx 2E_{\rm sym}(\rho_0)$ which becomes exact when both the kinetic and potential symmetry energies are proportional to $(\rho/\rho_0)^{2/3}$. We notice that an outstanding challenge for the community is to better quantify the uncertainties involved in extracting the constraints on the $E_{\rm sym}(\rho)$ and $L(\rho)$. Methods for addressing this challenge have been put forward, see, e.g., refs. \cite{Witek,Jorge}. Moreover, serious efforts and steady progress in this direction are being made, see, e.g., ref. \cite{Jun16}.  

Of course, the inferred values of $E_{sym}(\rho_0)$ and $L(\rho_0)$ are always sensitive to the particular models used to extract them from the data.
The sizes of neutron-skins of heavy nuclei and the radii of neutron stars are among the most well-known and extensively studied observables. For example, systematic studies~\cite{Roc11} based on the Skyrme/Gogny Hartree-Fock (HF) and Relativistic Mean-Fielf (RMF) models using various interactions indicate that the size of neutron skin $R_{skin}\equiv\Delta r_{np}$ in $^{208}$Pb and the value of $L(\rho_0)$ are linearly related according to $\Delta r_{np} ({\rm fm})= 0.101 + 0.00147\cdot L(\rho_0) ({\rm MeV}) $.  Indeed, many of the analyses surveyed in refs. \cite{ste14,Li-han,Oer16} have used the measured neutron-skins to constrain the value of $L(\rho_0)$. In view of the ongoing experimental efforts to measure more accurately the sizes of neutron-skins of various nuclei, such as the PREX-II and CREX experiments at Jefferson National Laboratory \cite{Chuck-nskin}, we emphasize that the significance of knowing precisely neutron-skins of neutron-rich nuclei is much more than determining the \es0 and $L(\rho_0)$. Actually, the $\Delta r_{np}$ is much more sensitive to the $L(\rho)$ at sub-saturation densities than at $\rho_0$. Moreover, the $\Delta r_{np}$ is correlated with the size of proton-skins in momentum space (due to the isospin dependence of short-range correlations, protons are predicted to move faster than neutrons in neutron-skins of heavy nuclei) according to the Liouville's theorem and Heisenberg's uncertainty principle \cite{Cai-skin}. 

\begin{figure}[ht]
\includegraphics[width=6.5cm]{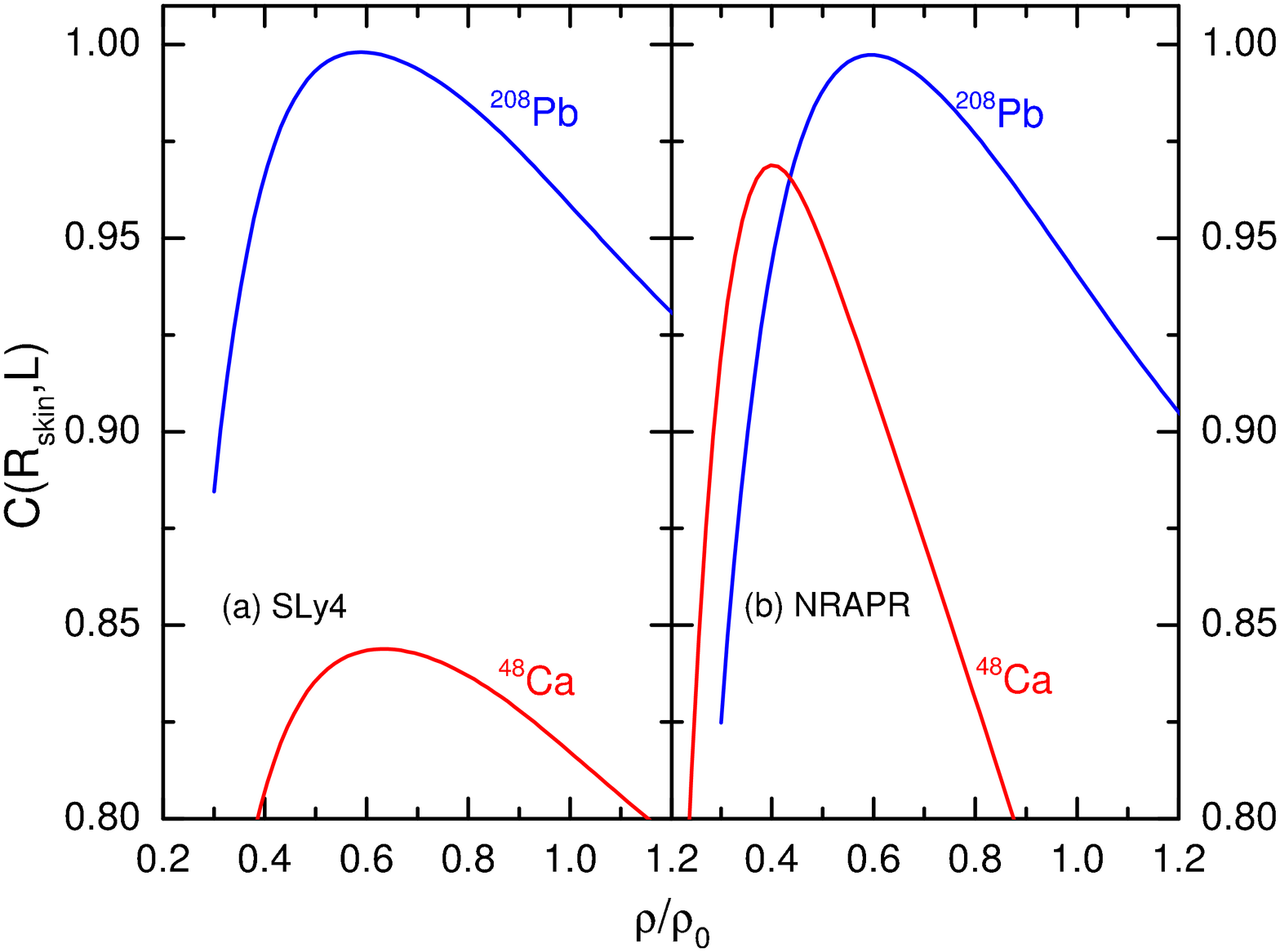}
\caption{Pearson's correlation coefficients between the $L(\rho)$ and the neutron skin thicknesses of $^{208}$Pb (blue) and $^{48}$Ca (red) as functions of density calculated
using SLy4 (left) and NRAPR (right) Skyrme energy density functionals. Taken from ref. \cite{Farrooh-skin}.}
\end{figure}
\begin{figure}[ht]
\smallskip
\includegraphics[width=6.5cm,angle=0]{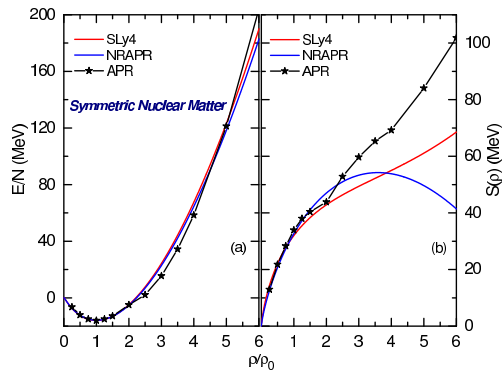}
\caption{Binding energy per nucleon in SNM (a) and the symmetry energy (b) as functions of the reduced density $\rho/\rho_0$ for SLy4 and NRAPR Skyrme energy density functionals. Taken from ref. \cite{Farrooh-skin}.} 
\end{figure}
\begin{figure}[ht]
\smallskip
\includegraphics[width=6.5cm,angle=0]{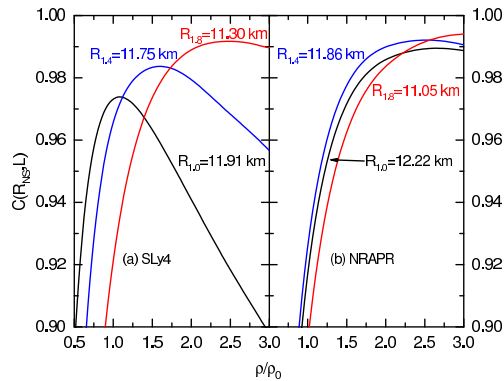}
 \caption{Pearson's correlation coefficients between $L(\rho)$ and the $1.0$-, $1.4$-, and $1.8$-solar mass neutron star radii as a
function of density calculated using SLy4 (left) and NRAPR (right) Skyrme energy density functionals with their default parameter sets. Taken from ref. \cite{Farrooh-skin}.} 
\end{figure}
As an example, shown in Fig. 4 is the Pearson's correlation coefficients between the $L(\rho)$ and the neutron skin thicknesses of $^{208}$Pb and $^{48}$Ca as functions of the reduced density $\rho/\rho_0$ using the SLy4 (left) and NRAPR (right) Skyrme energy density functionals. The results were obtained by allowing the isovector effective mass $m_{\rm v}^{\ast}(\rho_0)$ and the symmetry-gradient coefficient $G_{\rm v}$ to have a 20\% theoretical error-bars while fixing all isoscalar parameters at their default values~\cite{Farrooh-skin}. The two interactions with the default values of their parameters lead to almost identical EOS for SNM and \esym for densities up to about $1.5\rho_0$ \cite{Farrooh-skin} as shown in Fig. 5. As higher densities reached in the cores of neutron stars, the two interactions lead to quite different symmetry energies. It is seen from Fig. 4 that the strongest correlation coefficient appears only at a density of about $\rho/\rho_0 = 0.59$ for $^{208}$Pb and a slightly lower density for $^{48}$Ca with both interactions.  This confirms clearly the finding that nuclear observables related to average properties of nuclei constrains mostly the \esym at the so-called crossing or reference density, localised close to the mean value of the density of heavy nuclei: $\rho\approx 0.11$ fm$^{-3}$ \cite{Cen09,Khan12,Brown13,Mliu,Zhang-skin}. 

Similarly for the radii of neutron stars, although they have been used to constrain the \es0 and $L(\rho_0)$, they are much more sensitive to the behavior of \esym around $\rho_0$ to $3\rho_0$ rather than 
the $L(\rho_0)$ \cite{Lat01,Farrooh-skin}. Illustrated in Fig. 6 is the Pearson's correlation coefficient between the neutron star radii and the $L(\rho)$ as a function of density for both SLy4 and NRAPR models. In the case of SLy4, the radius of a 1.0-solar mass neutron star shows a strong correlation with the $L(\rho_0)$. For heavier neutron stars, the strongest correlation shifts to the $L(\rho)$ at higher densities, 
{\sl e.g.} at $1.5\rho_0$ for a 1.4-solar mass neutron star, and at $2.5\rho_0$ for a 1.8-solar mass neutron star. Moreover, the correlation coefficient remains almost
flat for higher densities in a 1.8-solar mass neutron star \cite{Farrooh-skin}. The difference in the high density behavior of the \esym using the  SLy4 and NRAPR interactions shows up in the correlation coefficient for more massive neutron stars. It is seen that the NRAPR exhibits a different evolution of the correlation coefficient with mass, with a much less pronounced increase in its peak towards higher masses because the underlying \esym softens, i.e., the \lr start decreasing with increasing density above about $1.5\rho_0$ \cite{Farrooh-skin}.\\

\begin{figure}[htb]
\includegraphics[width=12cm,height=6cm]{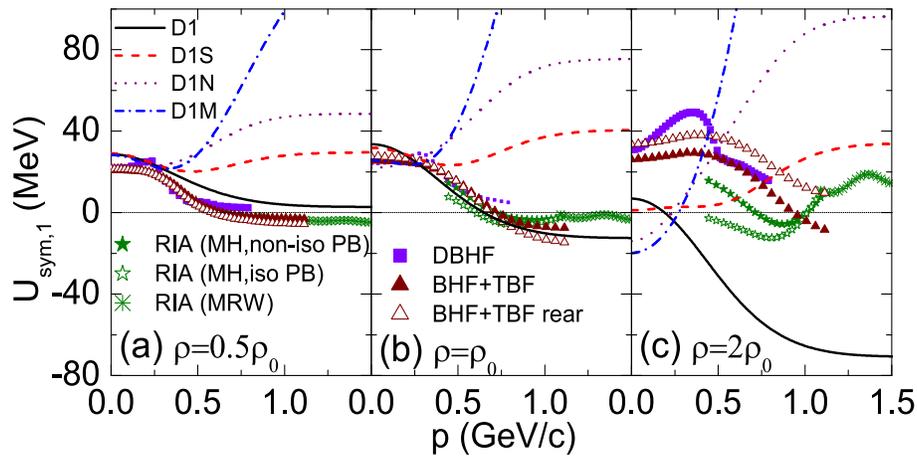}
\caption{Density and momentum dependence of the nucleon isovector potential predicted by the Gogny-Hartree-Fock calculations using the D1, D1S, D1M and D1N interactions, Dirac-Brueckner-Hartree-Fock (DBHF) and Relativistic Impulse Approximation (RIA) with various two-body and three-body forces (TBF). Taken from ref. \protect\cite{Rchen}.}\label{Usym1}
\end{figure}

\noindent{\bf Density dependence of nuclear symmetry energy}\\
While the community has made significant advancement in constraining the $E_{sym}(\rho_0)$ and $L(\rho_0)$, determining the density dependence of nuclear symmetry \esym and $L(\rho)$ away from the saturation density   are more challenging. First of all, model predictions are more diverse especially at high densities where the poorly known three-body force and possibly new degrees of freedom become important. The density and momentum dependence of the underlying isovector potential determining the \esym is very model dependent. As an example, shown in Fig.\ \ref{Usym1} are the predicted isovector potentials using the Gogny-Hartree-Fock, Dirac-Brueckner-Hartree-Fock and Relativistic Impulse Approximation with various two-body and three-body  interactions \cite{Rchen}. They all basically agree with the isovector potential at saturation density extracted from the optical model analyses of nucleon-nucleus scattering data \cite{Rchen,LiXH14}. Away from the saturation density, however, while some models predict decreasing symmetry potentials albeit at different rates, others predict instead increasing ones with growing nucleon momentum especially at high densities. The resulting \esym is then very different especially at supra-saturation densities. 

\begin{figure}[htb]
\includegraphics[clip, trim=3.5cm 0cm 0.0cm 1cm, width=12cm,height=7.cm]{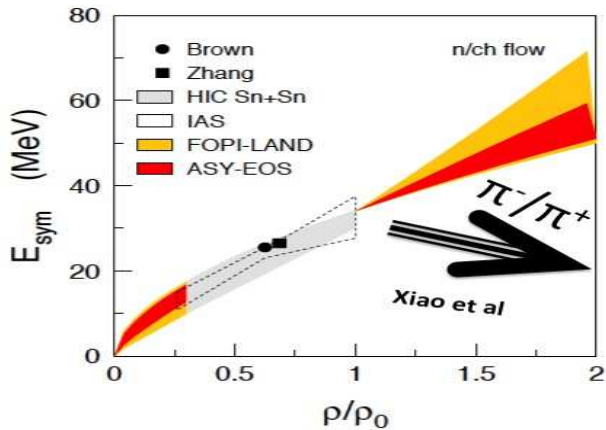}
\caption{{\protect Constraints on the density dependence of \esym using isospin diffusion data from MSU \cite{Tsang-PRL}, world data of excitation energies of the isobaric analog states (IAS) \cite{Pawel14}, isospin-dependent flow measurements by the ASY-EOS Collaboration at GSI \cite{russ11,ASY-EOS} in comparison with the trend (arrow) of \esym from an earlier analysis of the FOPI/GSI pion data by Xiao et al. using the IBUU04 transport model \cite{XiaoPRL} as well as the constraints at $\rho_0$ from analyzing properties of double magic nuclei by Brown \cite{Brown13} and binding energies and neutron-skins by Chen and Zhang \cite{Zhang-skin}. Taken from \cite{ASY-EOS}.}}\label{ASY}
\end{figure}

To experimentally probe the density dependence of symmetry energy, one needs to study systematics of static properties of nuclei or dynamical observables describing collective motions of nuclei or in nuclear reactions. We note that the observables used in constraining the $E_{\rm sym}(\rho_0)$ and $L(\rho_0)$ may also provide invaluable information about the \esym and \lr away from $\rho_0$. For example, it is well known that in the nuclear mass/energy formula of finite nuclei, the isospin asymmetry appears in both the volume and surface terms. Rewriting the nuclear contributions to the energy of finite nuclei of mass number A in a form similar to the Eq.1, i.e., $E(N,Z)=E_0(A)+a_{asy}(A)\cdot (N-Z)^2/A$ where $E_0(A)=-a_vA+a_sA^{2/3}$ is the symmetric part of the energy in terms of the volume and surface energy coefficients $a_v$ and $a_s$, 
one can define the mass dependence of the symmetry energy coefficient $a_{asy}(A)\equiv 1/[1/a^v_{asy}+A^{-1/3}/a^s_{asy}]$ in terms of the volume and surface symmetry energy coefficients $a^v_{asy}$ and $a^s_{asy}$ \cite{Pawel14}. A slightly different expression for the $a_{asy}(A)$ has been given in terms of the so-called neutron-skin stiffness coefficient $Q$ \cite{Roc11,Cen09,Zhang-skin}. The $a_{asy}(A)$ can be extracted from analyzing atomic masses \cite{Roc11,Cen09,Zhang-skin,Mliu} and/or excitation energies of the isobaric analog states \cite{Pawel14}. Supporting the proposition that $a_{asy}(A)=E_{sym}(\rho_A)$ at a reference density $\rho_A$ made in several studies \cite{Roc11,Cen09,Zhang-skin,Mliu}, 
the Pearson's correlation coefficient for $a_{asy}(A)$ based on SHF calculations does show a small bump at some  $\rho_A$ for a given A, e.g., $\rho_A\approx 0.11$ fm$^{-3}$ at A=240. However, the correlation coefficient remains rather large all the way down to zero density \cite{Pawel14}. By fitting the $a_{asy}(A)$ extracted from the isobaric states with SHF calculations, a constraining band on the \esym between approximately $\rho_0/3$ and $\rho_0$ was obtained by Danielewicz and Lee as shown in Fig. \ref{ASY}. Using the dipole polarizability data in a similar analysis \cite{Zhang-skin} results in a constraint consistent with the one from analyzing the IAS.  

Many reaction observables and phenomena from cross sections of sub-barrier fusion and fission at low energies \cite{fusion,YGMA}, energy and strength of various collective modes, isospin diffusion, isoscaling, hard photon production, ratios and differential flows of protons and neutrons as well as  mirror nuclei at intermediate energies, to pion, kaon and $\eta$ production in heavy-ion collisions up to 10 GeV/nucleon have been proposed as probes of the density dependence of nuclear symmetry energy, see, e.g., reviews collected in ref. \cite{Tesym}. Most of these observables probe directly either the $U_{\rm sym,1}(\rho,k)$ or consequences of the isospin fractionation. The latter refers to the phenomenon that in any density region where the symmetry energy is low/high, it is energetically more favorable for the region to be more/less isospin asymmetric at chemical equilibrium. In addition, for isovector collective modes of excited nuclei, the symmetry energy/potential plays the role of the restoring force \cite{Jorge2,Colo1,Colo2}. Since the isovector potential is normally small compared to the isoscalar potential, isospin-sensitive observables often involve relative or differential quantities/motions of neutrons and protons to enhance (reduce) effects of the isovector (isoscalar) potential \cite{LiBA96,LiBA97a,LiBA00,Chen05a}.  
Depending on the conditions of the reactions, these observables often probe the \esym over a broad density range. To extract the \esym from nuclear reactions one often uses dynamical models, such as transport models for heavy-ion collisions \cite{ireview98,ibook01,ditoro,LCK08}. 

Among many interesting experiments, it is worth emphasizing that significant work has been done in constraining the \esym using heavy-ion reactions at intermediate energies, see, e.g., refs. \cite{Lynch09,Trau12,Tsang12,Hag14,Sherry14,Fil14,Ade14,Khoa,Alan,Hud14} for recent reviews. For example, several transport model analyses of the experimental data on isospin diffusion between several Sn isotopes taken 
by Tsang et al. \cite{Tsang-PRL} at MSU have consistently extracted a constraining band on the \esym between approximately $\rho_0/3$ and $\rho_0$ shown in Fig. \ref{ASY}.  While at supra-saturation densities the data is very limited and transport model calculations of reaction observables are still rather model dependent \cite{Feng10,cozma11,Guo13,Yong16}. For example, analyzing the $\pi^-/\pi^+$ data from the FOPI/GSI collaboration \cite{FOPI,XiaoPRL} using a BUU-type (Boltzmann-Uehling-Uhlenbeck) transport model \cite{LiBA04a}, the \esym was found to decrease with increasing density above about $2\rho_0$ as predicted by the Gogny-Hartree-Fock calculations \cite{Das03}. Later, the ASY-EOS Collaboration analyzed the relative flows of neutrons w.r.t. protons, tritons w.r.t. $^3$He and yield ratios of light isobars using two versions of the QMD-type (Quantum Molecular Dynamics) transport models \cite{russ11,ASY-EOS}. As shown in Fig.\ \ref{ASY}, there is a clear disagreement regarding the high density behavior of nuclear symmetry energy based on these model analyses of the available data. Ironically, this is very similar to the current situation of microscopic many-body predictions for the high density \esym using various three-body forces with or without considering effects of the tensor force, see, e.g., ref. \cite{mach16}. Certainly, ongoing and planned new experiments coupled with better coordinated theoretical efforts using systematically tested reaction models will help improve the situation hopefully in the near future.\\

\noindent{\bf Concluding remarks and outlook}\\
This article provides my personal views on the physics motivations, community achievements and a few open issues in extracting the density dependence of nuclear symmetry energy from laboratory experiments. The views presented might be biased, incomplete and different from opinions of some other physicists working in this field. But I am sure we agree that the density dependence of nuclear symmetry energy \esym is an important quantity relevant for many interesting issues in both nuclear physics and astrophysics. Its determination has broad impacts. Besides the challenges in treating nuclear many-body problems, our poor knowledge about the isovector nuclear interaction is at the origin of the uncertain density dependence of nuclear symmetry energy. At the mean-field level, the density and momentum dependence of both the isoscalar and isovector single-nucleon potentials affects the \esym and its density slope $L(\rho)$. Going beyond the mean-field level, correlations and fluctuations, especially the short-range correlation due to the tensor force in the neutron-proton isosinglect channel also affects the symmetry energy especially at supra-saturation densities. Besides possible phase transitions, the high density symmetry energy has been the most uncertain term of the EOS of neutron-rich nucleonic matter. 
Thanks to the hard work of many people in both nuclear physics and astrophysics, much progress has been made in constraining the symmetry energy around and below the saturation density. In particular, rather consistent values of \es0$\sim 32$ MeV  and $L(\rho_0)\sim 59$ MeV have been obtained from over 50 analyses using various kinds of data and models, indicating a density dependence of the symmetry energy $E_{sym}(\rho)/E_{sym}(\rho_0)\approx (\rho/\rho_0)^{2/3}$ around the saturation density. However, the uncertainties of some of these analyses need to be better quantified. Moreover, the \esym at supra-saturation densities remains rather unconstrained. 

Looking forward, it is exciting to note that while some new experiments have been carried out recently or planned using dedicated new detectors, some dedicated working groups of theorists have self-organized themselves to address some of the key issues regarding the EOS of dense neutron-rich matter. Moreover, new opportunities are also being offered by the advanced radioactive beam facilities permitting reactions with higher isospin-asymmetries, thus enlarging the observable effects induced by the isovector nuclear interaction. In addition, new experiments using electron-nucleus and (p,2pN) reactions at large momentum transfers investigating the isospin dependence of short-range correlations in neutron-rich nuclei are being carried out or planned to better understand effects of the tensor force. On the other hand, new astrophysical observations, most noticeably the radii, the frequencies of torsional oscillations, the r-mode instability window of neutron stars, the neutrino flux from supernovae explosions, the cooling curves of protoneutron stars, the gravitational waves from collisions involving neutron stars, etc, also provide exciting new opportunities for better constraining the density dependence of nuclear symmetry energy, see, e.g., refs.\cite{Steiner05,Lat13,Blaschke16,ste14,Lat01,Newton14,Iida14,Pearson,Farr14,Fis14,Ozel}. These studies are being carried out or planned by using X-ray data from missions such as Chandra, XMM-Newton and the Neutron Star Interior Composition Explorer as well as gravitational waves from advanced detectors both on earth and in space. Combining new information from both terrestrial nuclear experiments and astrophysics observations will certainly allow us to determine much more precisely the symmetry energy of dense neutron-rich nucleonic matter. \\

\noindent{\bf Acknowledgement}\\
I would like to thank Wolfgang Bauer, Ignazio Bombaci, Bao-Jun Cai, Lie-Wen Chen, Maria Colonna, Champak B. Das, Pawel Danielewicz, Charles Gale, Umesh Garg, Vincenzo Greco, Francesca Gulminelli, Subal Das Gupta, Chuck Horowitz, Hyun Kyu Lee, Dao Tien Khoa, Massimo Di Toro, Farrooh Fattoyev, William Lynch, Yvonne Leifels, Wenjun Guo, Xiao-Tao He, Or Hen, Plamen Krastev, Wei-Zhou Jiang, Che Ming Ko, James Lattimer, Ang Li, Xiao-Hua Li, Yu-Gang Ma, Joseph Natowitz, William G. Newton, Akira Ono, Li Ou, Eli Piasetzky, Jorge Piekarewicz, Willy Reisdorf, Mannque Rho,  Tusar Ranjan Routray, W. Udo Schr\"{o}der, Andrew Steiner, Jirina Stone, 
Andrew Sustich, Stefan Typel, Wolfgang Trautmann, Betty Tsang, Isaac Vidana, Giusseppe Verde, De-Hua Wen, Larry B. Weinstein, Hermann Wolter, Chang Xu, Jun Xu, Zhi-Gang Xiao, Sherry Yennello, Gao-Chan Yong, Feng-Shou Zhang and Wei Zuo for collaborations and/or helpful discussions over the years on many of the issues discussed in this article. I would also like to thank Akira Ohnishi for inviting me to attend the YIPQS long-term and Nishinomiya-Yukawa memorial workshop on Nuclear Physics, Compact Stars, and Compact Star Mergers 2016 (YITP-T-16-02) at Kyoto University where I enjoyed interesting discussions with many other participants and started preparing this article. This work is supported in part by the U.S. Department of Energy, Office of Science, under Award Number DE-SC0013702 and the National Natural Science Foundation of China under Grant No. 11320101004.


\end{document}